\documentclass[conference]{IEEEtran}
\usepackage[printonlyused]{acronym}
\usepackage[style=ieee,backend=bibtex]{biblatex}
\bibliography{biblio}
\IEEEoverridecommandlockouts
\usepackage{amsmath,amssymb,amsfonts}
\usepackage{algorithmic}
\usepackage{graphicx}
\usepackage{array}
\usepackage{multirow}
\usepackage{booktabs}
\usepackage{textcomp}
\usepackage{xcolor}
\usepackage{mathtools}
\usepackage{placeins}
\usepackage{afterpage}
\usepackage{hyperref}
\usepackage{multicol}
\usepackage{bm}
\usepackage{enumitem}
\usepackage{etoolbox}
\makeatletter
\patchcmd{\@makecaption}
  {\scshape}
  {}
  {}
  {}
\makeatletter
\patchcmd{\@makecaption}
  {\\}
  {.\ }
  {}
  {}
\makeatother
\def\tablename{TABLE}

\def\BibTeX{{\rm B\kern-.05em{\sc i\kern-.025em b}\kern-.08em
    T\kern-.1667em\lower.7ex\hbox{E}\kern-.125emX}}

\makeatletter
\renewcommand{\fnum@figure}{Figure \thefigure}
\makeatother

\begin{document}

\title{Analytical Modelling of Raw Data for Flow-Guided In-body Nanoscale Localization}
\vspace{-7mm}
\author{
\IEEEauthorblockN{Guillem Pascual\IEEEauthorrefmark{1}, Filip Lemic\IEEEauthorrefmark{1}\textsuperscript{\textsection}, Carmen Delgado\IEEEauthorrefmark{1}, Xavier Costa-P\'erez\IEEEauthorrefmark{1}\IEEEauthorrefmark{2}}
\vspace{1mm}
\IEEEauthorblockA{\IEEEauthorrefmark{1}AI-Driven Systems, i2CAT Foundation, Spain}
\IEEEauthorblockA{\IEEEauthorrefmark{2}NEC Laboratories Europe GmbH, Germany and ICREA, Spain\\
Email: \{name.surname\}@i2cat.net}
\vspace{-9mm}
}

\maketitle

\begingroup\renewcommand\thefootnote{\textsection}
\footnotetext{Corresponding Author.}
\endgroup

\begin{abstract}
Advancements in nanotechnology and material science are paving the way toward nanoscale devices that combine sensing, computing, data and energy storage, and wireless communication. 
In precision medicine, these nanodevices show promise for disease diagnostics, treatment, and monitoring from within the patients' bloodstreams.
Assigning the location of a sensed biological event with the event itself, which is the main proposition of flow-guided in-body nanoscale localization, would be immensely beneficial from the perspective of precision medicine.
The nanoscale nature of the nanodevices and the challenging environment that the bloodstream represents, result in current flow-guided localization approaches being constrained in their communication and energy-related capabilities.
The communication and energy constraints of the nanodevices result in different features of raw data for flow-guided localization, in turn affecting its performance.
An analytical modeling of the effects of imperfect communication and constrained energy causing intermittent operation of the nanodevices on the raw data produced by the nanodevices would be beneficial.
Hence, we propose an analytical model of raw data for flow-guided localization, where the raw data is modeled as a function of communication and energy-related capabilities of the nanodevice. 
We evaluate the model by comparing its output with the one obtained through the utilization of a simulator for objective evaluation of flow-guided localization, featuring comparably higher level of realism. 
Our results across a number of scenarios and heterogeneous performance metrics indicate high similarity between the model and simulator-generated raw datasets. 
\end{abstract}



\acrodef{ML}{Machine Learning}
\acrodef{THz}{Terahertz}
\acrodef{GNN}{Graph Neural Networks}
\acrodef{ZnO}{Zinc Oxide}
\acrodef{IMU}{Inertial Measurement Unit}
\acrodef{RF}{Radio Frequency}
\acrodef{SINR}{Signal to Interference and Noise Ratio}
\acrodef{NN}{Neural Network}
\acrodef{HCS}{Human Cardiovascular System}
\acrodef{ECDF}{Empirical Cumulative Distribution Function}
\acrodef{KL}{Kullback-Leibler}
\acrodef{MW}{Mann-Whitney}
\acrodef{BVS}{BloodVoyagerS}
\acrodef{MSE}{Mean Squared Error}

\section{Introduction}

Contemporary advancements in nanotechnology are creating an opportunity for the development of nanoscale devices that combine sensing, computing, and data and energy storage functionalities~\cite{jornet2012joint}. 
These nanodevices are envisaged to revolutionize a variety of applications in precision medicine~\cite{abbasi2016nano}. 
Some applications involve deploying nanodevices in the patients' bloodstreams, requiring their physical size to be comparable to that of the red blood cells (i.e., up to 5 microns). 
Due to their minuscule dimensions, these nanodevices will rely on nanoscale components for harvesting energy from environmental sources such as heartbeats or ultrasound power transfer~\cite{jornet2012joint}. 
As a consequence, these devices are anticipated to passively circulate within the bloodstream.

Advancements in advanced materials, notably graphene and its derivatives~\cite{abadal2015time}, have created possibilities for nanoscale wireless communication at \ac{THz} frequencies (i.e., 0.1-10 THz)~\cite{lemic2021survey}. 
The inclusion of wireless communication capabilities enables two-way communication between nanodevices and the external world~\cite{dressler2015connecting}. 
Nanodevices with integrated communication abilities are facilitating sensing-based applications such as oxygen sensing in the bloodstream for early cancer diagnosis and actuation-based applications like non-invasive targeted drug delivery for cancer treatment. 
Additionally, nanodevices with communication capabilities serve as a foundation for flow-guided localization within the bloodstream ~\cite{lemic2021survey}. 
Flow-guided in-body localization could enable the association of a physical location to an event detected by a nanodevice, offering advantages such as non-invasiveness, early and precise diagnostics, and cost reduction~\cite{lemic2022toward,gomez2022nanosensor,bartra2023graph}.

Localization methods from~\cite{bartra2023graph,gomez2022nanosensor} operate in a way that, in each passage through the heart, the nanodevices try to establish communication with an on-body anchor for transmitting their identifiers along with a binary indicator of whether an event was detected. 
The anchor utilizes this to calculate the elapsed time since the previous transmission, enabling the inference of a cardiovascular path that contains the detected event. 
At \ac{THz} frequencies, challenges arise due to molecular absorption in the medium, leading to attenuation, distortion, and additional noise~\cite{dressler2015connecting}. 
As a consequence, the ability of nanodevices to establish effective communication with the anchor is affected, as indicated in e.g.,~\cite{lopez2023toward,lemic2023insights}.
This in turn results in erroneous data being eventually transmitted to the anchor, as the transmitted data would encapsulate multiple iterations through the bloodstream.
Note that other nanocommunication approaches potentially suitable for in-body nanocommunication (e.g., molecular, magnetic, or ultrasound), are expected to feature similar communication unreliability due to e.g., significant path loss, high mobility, and energy or size constrains at the nanodevice level, as discussed in e.g.,~\cite{stelzner2017function}.

Localization systems from~\cite{bartra2023graph,gomez2022nanosensor,lemic2022toward} leverage nanocapacitors' piezoelectric effect of \ac{ZnO} nanowires. 
Despite its remarkable utility, this energy harvesting mechanism can result in the intermittent operation of harvesting entities~\cite{lemic2019modeling}. 
This process is usually dependent on a turn-off energy threshold that determines whether the nanodevice is activated and capable of detection.
Consequently, the energy constraints impact the nanodevice's ability to detect events, therefore impeding its capacity to transmit correct information to the anchor.

In summary, it is well-known that flow-guided localization accuracy is strongly dependent on the transmission and energy-related capabilities at the nanodevice level~\cite{lopez2023toward,lemic2023insights}. 
The main reasons can be found in the fact that these processes introduce erroneous raw data for flow-guided localization in the form of either compound iteration times or erroneous event detection indicators.
Hence, there is a need for modelling of the raw data for capturing the effects of these two sources of stochasticism in flow-guided localization. 
Toward addressing this issue, we propose an analytical model of raw data for flow-guided in-body nanoscale localization. 
The model encapsulated nanodevice mobility, in-body communication, and energy constraints of current flow-guided localization systems. 

We assess the utility of the proposed model by comparing its raw data output with the data generated under comparable conditions utilizing a state-of-the-art simulator for objective performance evaluation of flow-guided localization~\cite{lopez2023toward}.
For a variety of relevant evaluation scenarios and heterogeneous performance metrics, we demonstrate that the raw data generated through the model is highly comparable to the corresponding simulator-generated raw data with a significantly higher level of realism.


\section{Flow-guided In-Body Nanoscale Localization Overview and Framework}
\label{sec:model}

\begin{figure}[!t]
\centerline{\includegraphics[width=\columnwidth]{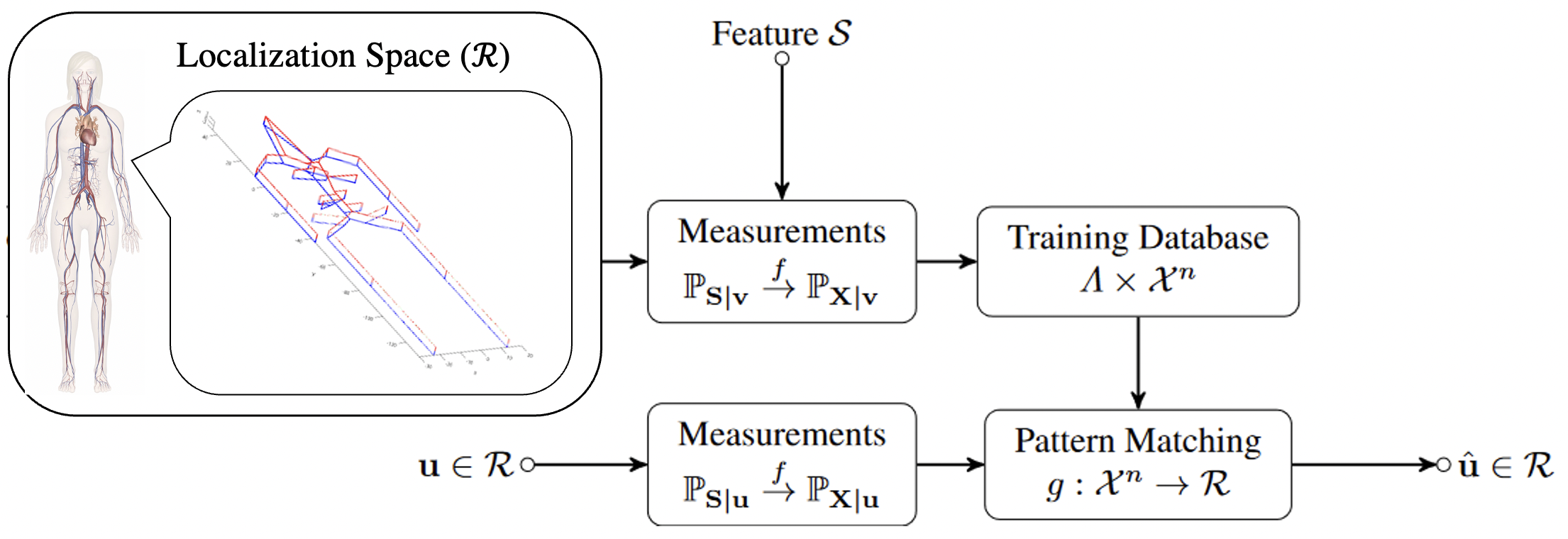}}
\caption{Flow-guided in-body nanoscale localization framework.}
\label{fig:diagrama}
\vspace{-3mm}
\end{figure}

Toward developing the analytical model, we consider a flow-guided localization framework and notations as depicted in Figure~\ref{fig:diagrama}.
The framework is adapted from~\cite{behboodi2016mathematical}, where the authors propose an analytical framework for more traditional fingerprinting-based indoor localization. 
Within this approach, a distinct feature of an environment is selected as the basis for creating the fingerprint. 
In the context of flow-guided localization, the environment is an entire bloodstream, modeled as a set of potential cardiovascular paths that a nanodevice might pass through in each of its iterations.

The utilized mathematical notations are summarized in Table~\ref{tab:symbols}.
The chosen signal feature is denoted as $S$ and belongs to the feature space $\mathcal{S}$. 
Consecutive observations of the signal feature, represented as $\bm{S} = (S_1, . . . , S_m) \in \mathcal{S}$, form a random vector that is linked to the location \(u\) through the conditional probability \(P_{S\mid u}\). 
Raw data for flow-guided localization corresponds to a fingerprint in fingerprinting-based one, and is constructed based on these observations, capturing the unique characteristics of the signal feature at each respective location.
The observed feature is then converted into raw data through a raw data-creating function. 

The subsequent stage involves the creation of a training database through measurements of the signal feature \(S\) at various training locations. This database serves as a reference for the subsequent location estimation. 
To determine the location of a nanodevice at \(u\), a pattern matching function \(g\) is utilized. 
By comparing the acquired raw data with the instances stored in the training database, the pattern matching function \(g\) estimates the location based on the closest matching instances. 
The summarized flow-guided localization framework is:

\begin{itemize}[leftmargin=*]
     \item \textbf{Localization space $\mathcal{R}$:} the cardiovascular system, possible detection regions \(\{R_1,....,R_r\}\), each with the iteration times \(\{T_1,...,T_r\}\).
     \item \textbf{Feature $\mathcal{S}$ = Raw data} ($\mathcal{X}$): Time between consecutive transmissions and event bit \((t,b)\).
     \item \textbf{Pattern matching function \(\bm{g}\):} \ac{ML}-based flow-guided localization algorithm.
\end{itemize}

Based on the outlined framework, our aim is to derive analytically the conditional probabilities \(P_{S\mid u}\). 
The parameters that will be used are:
\begin{itemize}[leftmargin=*]
    \item \textbf{Probability of detection \bm{$P_{det}$}} corresponds to the probability of an event being detected. This parameter encapsulates the intermittent nature of a nanodevice due to energy-harvesting, which might result in a nanodevice not detecting an event of interest as it was turned off, although it went through the path that contained the event. This results in an erroneous event bit $b$.
    \item \textbf{Probability of transmission \bm{$P_{trans}$}} corresponds to the probability of successfully transmitting data to the on-body anchor in the vicinity of the heart. It incorporates factors such as having sufficient energy for communication, being in the range of the anchor, and self-interference between nanodevices. If the data is not communicated properly to the anchor, the iteration time will not be reset, leading to compound iterations.
    
\end{itemize}

\begin{table}[!t]
    \centering
    \scriptsize
    \vspace{-1mm}
    \caption{Overview of utilized notations.}
    \vspace{-1mm}
    \label{tab:symbols}
        \begin{tabular}{c p{2.55cm} c p{3.1cm}}
        \hline
        \textbf{Symbol} & \textbf{Description} & \textbf{Symbol} & \textbf{Description} \\
        \hline
        $\mathcal{R}$ & Localization space  & $R_i$ & Region \\
        $\mathcal{S}$ & Feature space & $S$ & Feature \\
        $\mathcal{X}$ & Raw data space &  $X$ & Raw data \\
        $\Lambda$ & Training space & $w,u$ & Training/measurement location \\
        $g$ & Pattern matching function & $T_i$ & Iteration times \\    
        $P_{trans}$ & Transmission probability & $P_{det}$ & Detection probability \\
        \hline
        \end{tabular}
        \vspace{-3mm}
\end{table}


\section{Analytical Modelling of Raw Data for Flow-Guided In-body Nanoscale Localization}
\label{sec:prob_distrib}

Based on the notations established previously, raw data for flow-guided localization corresponds to a tuple \(X=(t,b)\) with \(t \in (0,M]\) and \(b \in \{0,1\}\). \(M\) represents the total duration that the nanodevices spend in the bloodstream.
The time elapsed between transmissions will be a composition of the travel times across different regions, accompanied by zero-mean distributed Gaussian noise \(Q\).
This noise factor accounts for different variations in iterations times due to factors such as turbulent blood-flow, short blood pressure variability, and similar biological factors. 
Hence, the iteration times can be modeled as a combination of deterministic travel times and random perturbations captured as the Gaussian noise.  
Thus, the raw data for flow-guided localization can be expressed as:

\vspace{-2mm}
\begin{equation}
X=\{(n_1T_1+...+n_rT_r+Q,b)\mid n_i \in \mathbb{N}, b \in \{0,1\}\},
\end{equation}
where $n_i$ represents the number of laps a nanodevice made through region $R_i$.
An important observation is that, depending on a region where the event is located, there exists a subset of raw data that can never occur. 
This subset corresponds to cases where the event has been detected ($b=1$) but the nanodevice has not passed though the region containing the event. Let
\(X_i\) represent the set of possible raw data instances for an event in region \(R_i\), then the mentioned subset must be substracted from the total set of fingerprints \(X\):

\vspace{-3mm}
\begin{equation}
X_i= X \backslash \{(n_1T_1+...+n_rT_r+Q,1)\mid n_i=0\}.
\end{equation}

We follow by modeling the sources of stohasticism in flow-guided localization:

\begin{itemize}[leftmargin=*]
    \item \textbf{Compound iterations:} As previously mentioned, if the nanodevice fails to communicate with the anchor, the iteration time will not reset, resulting in longer duration.
    \item  \textbf{False negatives:} When the nanodevice fails to detect the target, although it has passed through the affected region.
\end{itemize}

Given that we are dealing with a binary event bit $b$, we distinguish cases where the event is detected and vice-versa, and calculate the probabilities of interest as follows.

\subsection{Case 1: Event Detected ($b=1$)}
Suppose that the event to be detected is in region \(R_j\), leading to the following expressions:

\vspace{-3mm}
\begin{equation}
P(X=( n_1T_1+...+n_rT_r,1)\mid R_j)=P(\chi_{(n,1)}\mid R_j),
\end{equation}

where \(\chi_{(n,1)}\) refers to this particular raw data instance. 
There are multiple ways in which this raw data instance can be obtained taking into account that:
\begin{itemize}[leftmargin=*]
    \item There are multiple ways in which the nanodevice can travel this number of times through each cardiovascular path, this number is given by the number of permutations of the multiset \(\{P_{R_1}^{n_1},...,P_{R_r}^{n_r}\}\), where \(P_{R_i}\) is the probability of the nanodevice traveling through each region.
    The number of permutations corresponds to the expression:
    \vspace{-1mm}
    \begin{equation}
    \binom{n_1+...+n_r}{n_1,...,n_r}=\frac{(n_1+...+n_r)!}{n_1!...n_r!}.
    \end{equation}
    From that it follows:
    \vspace{-2mm}
    \begin{equation}
    P(\chi_{(n,1)}\mid R_j) = \binom{n_1+...+n_r}{n_1,...,n_r}P_{R_1}^{n_1}...P_{R_r}^{n_r}.
    \end{equation}

    \item The detection can occur in any iteration through \(R_j\) and once it is detected, the event bit will not change. 
    Let \(P_{d_1}\) be the probability of detecting the event in iteration \(i\), then:
    \vspace{-2mm}
    \begin{equation}
    P_{d_i}=(1-P_{det})^{i-1}P_{det}.
    \end{equation}
\end{itemize}

It is important to consider that the communication was only successful during the last iteration and not in the previous ones. 
Otherwise, the time would have been reset when the communication was successful. 
Thus, a multiplicative factor is applied to account for this condition, defined as follows.
\vspace{-1mm}
\begin{equation}
P_t=(1-P_{trans})^{(n_1+...+n_r-1)}P_{trans}.
\end{equation}
Finally, the total probability of having a certain raw data instance with event bit $b=1$ is:
\vspace{-1mm}
\begin{equation}
P(\chi_{(n,1)}\mid R_j)=\binom{n_1+...+n_r}{n_1,...,n_r}P_{R_1}^{n_1}...P_{R_r}^{n_r}P_t\sum_{i=1}^{n_j} P_{d_i}.
\end{equation}

\subsection{Case 2: Event not Detected ($b=0$)}

The probabilities for cases where no detection has occurred are the following, assuming an event located in region \(R_j\):
\vspace{-1mm}
\begin{equation}
P(X=( n_1T_1+...+n_rT_r,0)\mid R_j)=P(\chi_{(n,0)}\mid R_j).
\end{equation}

This expression is comparable to the former case, but here the event is not detected in any of the iterations through \(R_j\), which means that the following term will be always multiplying, this term will be defined as \(P_{nd}\), corresponding to the probabilities of not detecting an event in any iteration in which the event could have been detected:

\vspace{-2mm}
\begin{equation}
P_{nd}=(1-P_{det})^{n_j},
\end{equation}
leading to expression:
\vspace{-2mm}
\begin{equation}
P(\chi_{(n,0)}\mid R_j)=\binom{n_1+...+n_r}{n_1,...,n_r}P_{R_1}^{n_1}...P_{R_r}^{n_r}P_{nd}P_t.
\end{equation}

\subsection{An Example}

\begin{figure}[t]
\centerline{\includegraphics[width=0.97\columnwidth]{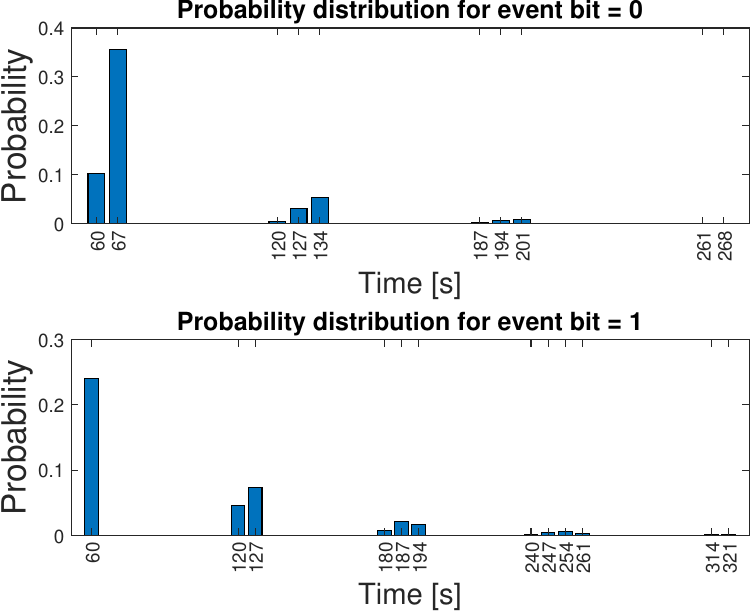}}
\vspace{-2mm}
\caption{Probability distribution for two arbitrary regions with different traveling times (60,67), with probabilities: \(P_{det}=0.7, P_{trans}=0.7\).}
\label{fig:Probd}
\vspace{-3mm}
\end{figure}

As an example, two arbitrary detection regions \((R_1,R_2)\) are considered, with  corresponding traveling times \((T_1,T_2)=(60,67)\)~[sec] and probabilities \((P_{R_1},P_{R_2})=(0.49,0.51)\). 
The probability distributions can be computed using the probabilities (\(P_{R_i}\)) and traveling times (\(T_i\)) of each region, the detection (\(P_{det}\)) and transmission (\(P_{trans}\)) probabilities, the region containing the target event, and the duration of the administration of the nanodevices in the bloodstream. 
From that, it uses a recursive algorithm to account for all possible combinations of iterations with different times, and from there it computes the probability for every case.
Figure~\ref{fig:Probd} shows the probability distribution assuming a target event in \(R_1\), with $P_{det}=0.7$ and $P_{trans}=0.7$. 
As visible, the highest probabilities correspond to the traveling times of each region, however there are also compound iterations with certain probabilities, for example the ones corresponding to 120, 127, and 134~s. 
The probabilities become practically unnoticeable after three compound iterations, suggesting the system convergence after several iterations trough the bloodstream.

\begin{figure}[t]
\centerline{\includegraphics[width=6cm]{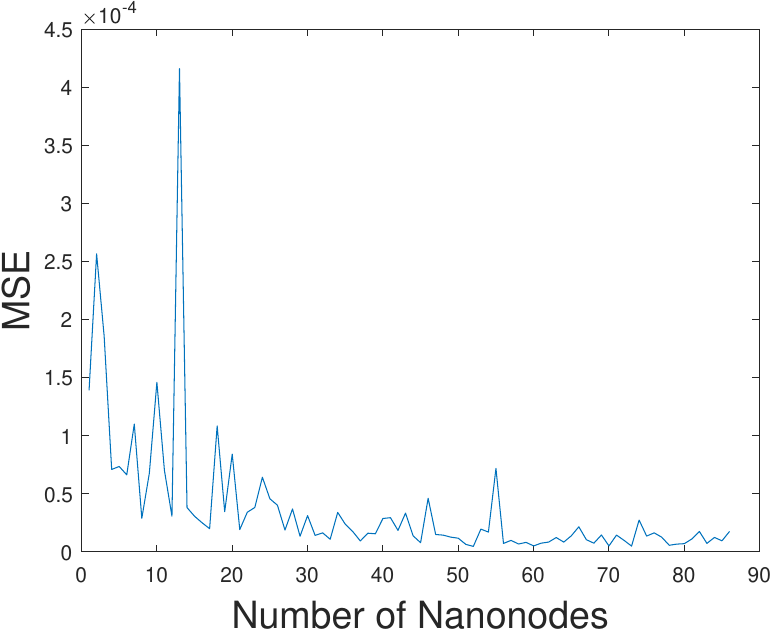}}
\vspace{-2mm}
\caption{MSE of generated frequencies with respect to the probability distribution as a function of nanonodes.}
\label{fig:mse}
\vspace{-4mm}
\end{figure}

Until now, we have considered as singe nanodevice deployed in the bloodstream, although a number of applications envision the administration of a large number of such devices. 
The presence of more nanodevices leads to an augmented volume of data, which can be assessed by plotting accuracy measures against the sample size.
It is anticipated that the frequencies will ultimately converge to the values of the probability distribution. 
To discern the rate of convergence, we can compute the \ac{MSE} between generated data frequencies and the probability distribution while varying the number of nanodevices.
Example results are depicted in Figure~\ref{fig:mse}, revealing that the \ac{MSE} decreases rapidly with the number of nanonodes, indicating that the distribution of raw data does not change significantly with an increase in the number of nanodevices, apart from increasing the frequency of data transmissions, indicating the utility of the proposed model for modelling the raw data obtained by simultaneously utilizing more than one nanodevice.

\section{Evaluation Setup and Results}
\label{sec:comparison}

\subsection{Evaluation Setup}

For assessing the accuracy of the proposed model of raw data for flow-guided in-body nanoscale localization, we will compare the data yielded by the model with the corresponding one generated by the state-of-the-art simulator for objective performance benchmarking of flow-guided localization~\cite{lopez2023toward}. 
The utilized simulator features significantly higher level of realism compared to the model.
Specifically, it combines \ac{BVS}~\cite{geyer2018bloodvoyagers} for modeling the mobility of the nanodevices in the bloodstream and ns-3-based TeraSim~\cite{hossain2018terasim} for \ac{THz}-based nanoscale communication between the nanodevices and the outside world, accounting for the energy-related and other technological constraints (e.g., pulse-based modulation) at the nanodevice level.

The \ac{BVS} simulator encompasses a comprehensive set of 94 vessels and organs, utilizing a coordinate system centered on the heart. 
All organs share an identical spatial depth, calibrated to a reference thickness of 4 cm, resembling the dimensions of a kidney. 
The simulator also assumes a predefined arrangement with arteries positioned anteriorly and veins posteriorly. 
The transitions between arteries and veins occurs within organs, limbs, and head, which jointly account for 24 regions in the body, as indicated in Table~\ref{tab:body_parts}. 
In the heart, the blood undergoes a transition from veins to arteries, signifying a shift from posterior to anterior flow. 
The simulator models the flow rate based on the relationship between pressure differences and flow resistance, yielding average blood velocities of 20~cm/sec in the aorta, 10~cm/sec in arteries, and 2-4~cm/sec in veins. 
Transitions between arteries and veins are simplified by assuming a constant velocity of 1~cm/sec.

TeraSim~\cite{hossain2018terasim} is the pioneering simulation platform tailored for modeling \ac{THz} (nano)communication networks. 
This platform accurately captures the unique capabilities of nanodevices and distinctive characteristics of THz signal propagation. 
TeraSim is integrated as a module within ns-3, a discrete-event network simulator, and it incorporates specialized physical and link layer solutions optimized for nanoscale THz communications. 
Specifically, at the physical layer, TeraSim implements pulse-based communication with omnidirectional antennas, catering to distances shorter than 1~m assuming a single transmission window of nearly 10~THz. 
At the link layer, TeraSim incorporates two well-established protocols, namely ALOHA and CSMA. 
Additionally, we have introduced in TeraSim a shared \ac{THz} channel module that implements a frequency-selective in-body wireless nanocommunication channel~\cite{gomez2023optimizing}.

Our analytical model has been instantiated on the 24 regions as modeled by the \ac{BVS}.
Other simulation parameters such as the transmit power, receiver sensitivity, number of nanonodes, operating frequency, and communication bandwidth have been consistently parameterized across the model and simulator. 
For each considered evaluation scenario, two sets of raw data are generated, one with a varying probability of transmission $P_{trans}$ and ideal detection probability (i.e., $P_{det}=1$), and vice-versa (i.e., $P_{trans}=1$). 
These probabilities have been hard-coded in the simulator to assess the consistency of the raw data outputs between the model and simulator.
Additionally, a realistic scenario is considered in which both probabilities are set to non-ideal values to assess the capabilities of the model in capturing their joint effects on the raw data.

The following metrics are employed to assess the performance similarity across the raw datasets: 
\paragraph{\textbf{\ac{MW} test}} will be used to evaluate the similarity between the distributions of iteration times between the model and simulator for body regions containing an event (i.e., for event bit $b=1$). 
\paragraph{\textbf{Square difference between \acp{ECDF}}} will be generated for event bit values $b=0$ for all body regions. 
The maximum vertical distance between \acp{ECDF} will be computed and averaged over all regions, and provided graphically utilizing regular box-plots, facilitating qualitative interpretation of the data.
\paragraph{\textbf{\ac{KL} divergence}} will be employed to compare the difference between ratios of ones and zeros in each region for varying transmission and detection probabilities. 
This approach helps identifying regions in which the fraction of event bits $b=1$ is comparable across the model and simulator. 
Regions through which the nanonodes pass in each of their iteration through the bloodstream (lungs and right heart) will be excluded from this metric. 

\begin{table}[!t]
    \centering
    \scriptsize
    \caption{BVS-based instantiation of the framework to 24 body regions.}
    \vspace{-3mm}
    \label{tab:body_parts}

    \begin{multicols}{3}

        \begin{tabular}{cl}
        \hline
        \textbf{ID} & \textbf{Body Part} \\
        \hline
        1 & Head \\
        2 & Thorax \\
        3 & Right shoulder \\
        4 & Left shoulder \\
        5 & Spleen \\
        6 & Right upper arm \\
        7 & Left upper arm \\
        8 & Liver \\
        \hline
        \end{tabular}        
        
        \begin{tabular}{cl}
        \hline
        \textbf{ID} & \textbf{Body Part} \\
        \hline
        9 & Right elbow \\    
        10 & Intestine \\
        11 & Right hand \\
        12 & Kidneys \\
        13 & Left elbow \\
        14 & Left hand \\
        15 & Right hip \\
        16 & Left hip \\
        \hline
        \end{tabular}

         \begin{tabular}{cl}
        \hline
        \textbf{ID} & \textbf{Body Part} \\
        \hline
        17 & Right knee \\     
        18 & Left pelvis \\
        19 & Left knee \\
        20 & Right pelvis \\
        21 & Right foot \\
        22 & Left foot \\
        23 & Lungs \\
        24 & Right heart \\
       \hline
        \end{tabular}
    \end{multicols}
    \vspace{-4mm}
\end{table}

\subsection{Evaluation Results}
\begin{table}[t]
  \centering
  \caption{Mann-Whitney test results for event bit $b=1$.}
  \vspace{-2mm}
    \label{table:mannwhitney}
  \begin{tabular}{cccc}
    
    \toprule
    \multirow{2}{*}{$P_{trans}$/$P_{det}$} & \multicolumn{2}{c}{Fraction of accepted Mann-Whitney tests.} \\
    \cmidrule{2-3}
    & $P_{det}=1$ & $P_{trans}=1$ \\
    \midrule
    0.2 & 83\% & 92\% \\
    0.4 & 88\% & 79\% \\
    0.6 & 94\% & 75\% \\
    0.8 & 83\% & 79\% \\
    \bottomrule
  \end{tabular}
\vspace{-4mm}
\end{table}

The \ac{MW} test results presented in Table~\ref{table:mannwhitney} demonstrate that the null hypothesis of both datasets being equally distributed is predominantly accepted across diverse regions, with a particularly strong performance observed for the ideal detection probability. 
Moreover, upon closer examination, a distinct pattern of higher variability emerges in the regions that do not meet the test's criteria, suggesting the potential influence of unspecified stochastic factors on the model's behavior in a small number of specific cases. 
Nonetheless, it is evident that the similarity between the model and the simulator distributions for event bit $b=1$ is notably high across the considered scenarios.

\begin{table}[t]
  \centering
  \caption{ECDF comparison for event bit $b=0$.}
  \vspace{-2mm}
  \label{table:ecdf}
  \begin{tabular}{cccc}
    \toprule
    \multirow{2}{*}{$P_{trans}$/$P_{det}$} & \multicolumn{2}{c}{Mean of vertical distance between ECDFs} \\
    \cmidrule{2-3}
    & $P_{det}=1$ & $P_{trans}=1$ \\
    \midrule
    0.2 & 0.11 & 0.081 \\
    0.4 & 0.12 & 0.084 \\
    0.6 & 0.1 & 0.083 \\
    0.8 & 0.1 & 0.08 \\
    \bottomrule
  \end{tabular}
  \vspace{-5mm}
\end{table}

\begin{figure}[!t]
\centerline{\includegraphics[width=0.9\columnwidth]{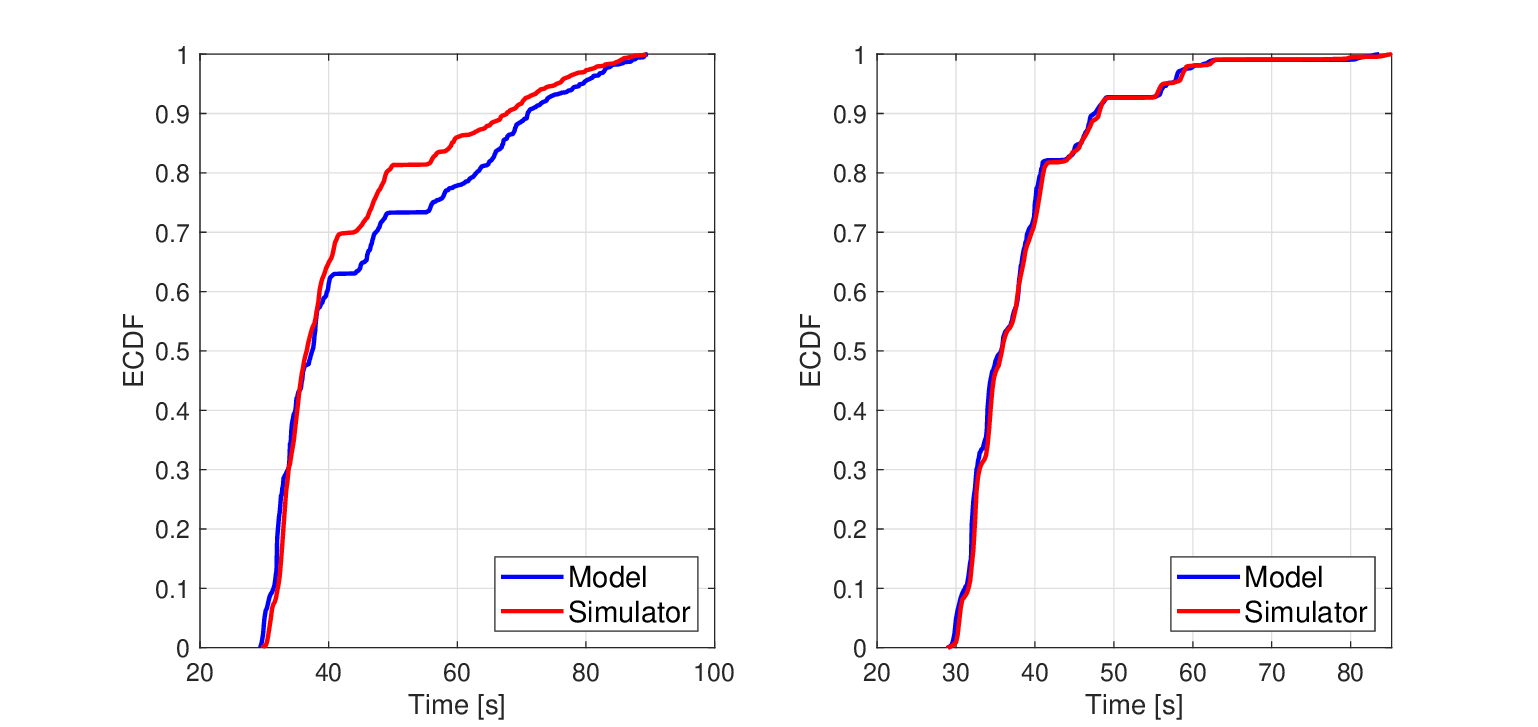}}
\vspace{-2mm}
\caption{ECDF comparison between model and simulator for $P_{trans}=0.4$ (left) and $P_{det}=0.4$ (right) in thorax.}
\label{fig:ecdfp}
\vspace{-2mm}
\end{figure}

For scenarios involving event bit $b=0$, insights into the performance similarity between the model and the simulator are assessed utilizing \acp{ECDF}. 
The average maximum vertical distances between \acp{ECDF} are shown in Table~\ref{table:ecdf}. 
Particularly, a higher level of similarity between ECDFs is observed for the transmission probability $P_{trans}=1$ compared to $P_{det}=1$. 
This difference can be attributed to the presence of compound iterations when varying the transmission parameter (cf., Figure~\ref{fig:Probd}). 
It is worth noting that this discrepancy predominantly surfaces within the distributions' midsection, with the tails being nearly indistinguishable. 
An illustrative instance of this behavior is showcased in Figure~\ref{fig:ecdfp}. 
In summary, despite notable differences in a small number of specific cases, the focal point of dissimilarity is confined to a specific region within the distribution. 
This implies that, in select instances, the distribution is subtly shifted in a defined segment while otherwise featuring a significant level of similarity.

\begin{figure}[!t]
\centerline{\includegraphics[width=\columnwidth]{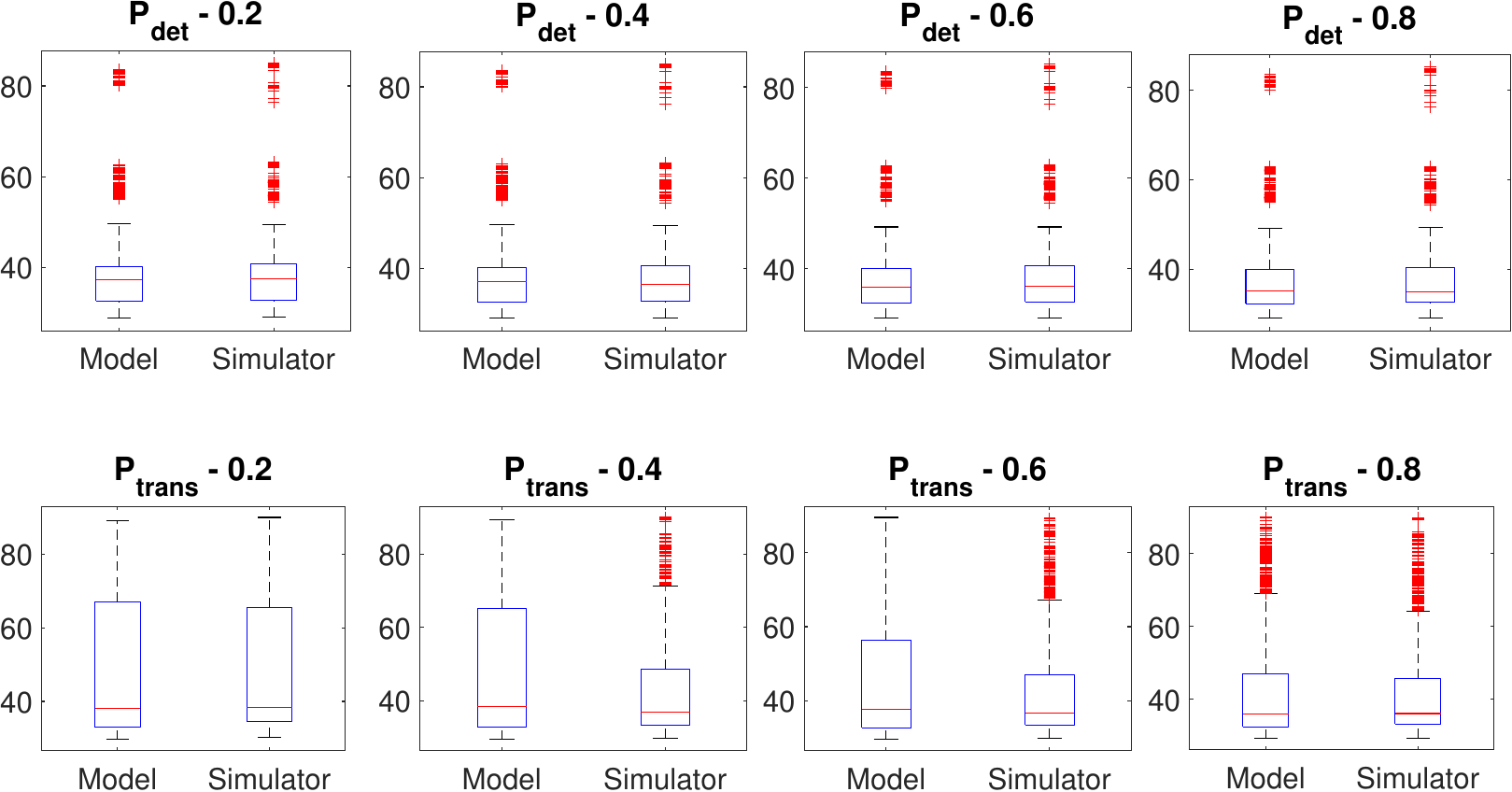}}
\vspace{-2mm}
\caption{Iteration times obtained with event bit $b=1$ for the head region.}
\label{fig:boxplots}
\vspace{-5mm}
\end{figure}

To further examine the distributions for event bit $b=0$, the distributions of iteration times in different scenarios and for the head region are shown in Figure~\ref{fig:boxplots}.
As visible, the distributions are shown to be highly comparable, apart in cases when $P_{trans}$ equals 0.4 and 0.6.  
Despite the differences in the distributions depicted for these two cases, an underlying similarity in the data distribution is evident. 
Specifically, while the simulator-generated boxplots exhibit a substantial number of outliers, in the model-related boxplots these outliers are considered within the third quartile. 
It is important to recognize that these outliers, although visually distinct, are not indicative of a fundamental divergence in the dataset's core characteristics. 
This can be checked in the case for 0.2 transmission probability where the simulator boxplots are almost equal to the ones originating from the model as the number of outliers is smaller, thus the third quartile is extended. 
These outliers are the root cause of the difference between \acp{ECDF}, as seen in Figure~\ref{fig:ecdfp}. 

\begin{figure}[t]
\centerline{\includegraphics[width=0.9\columnwidth]{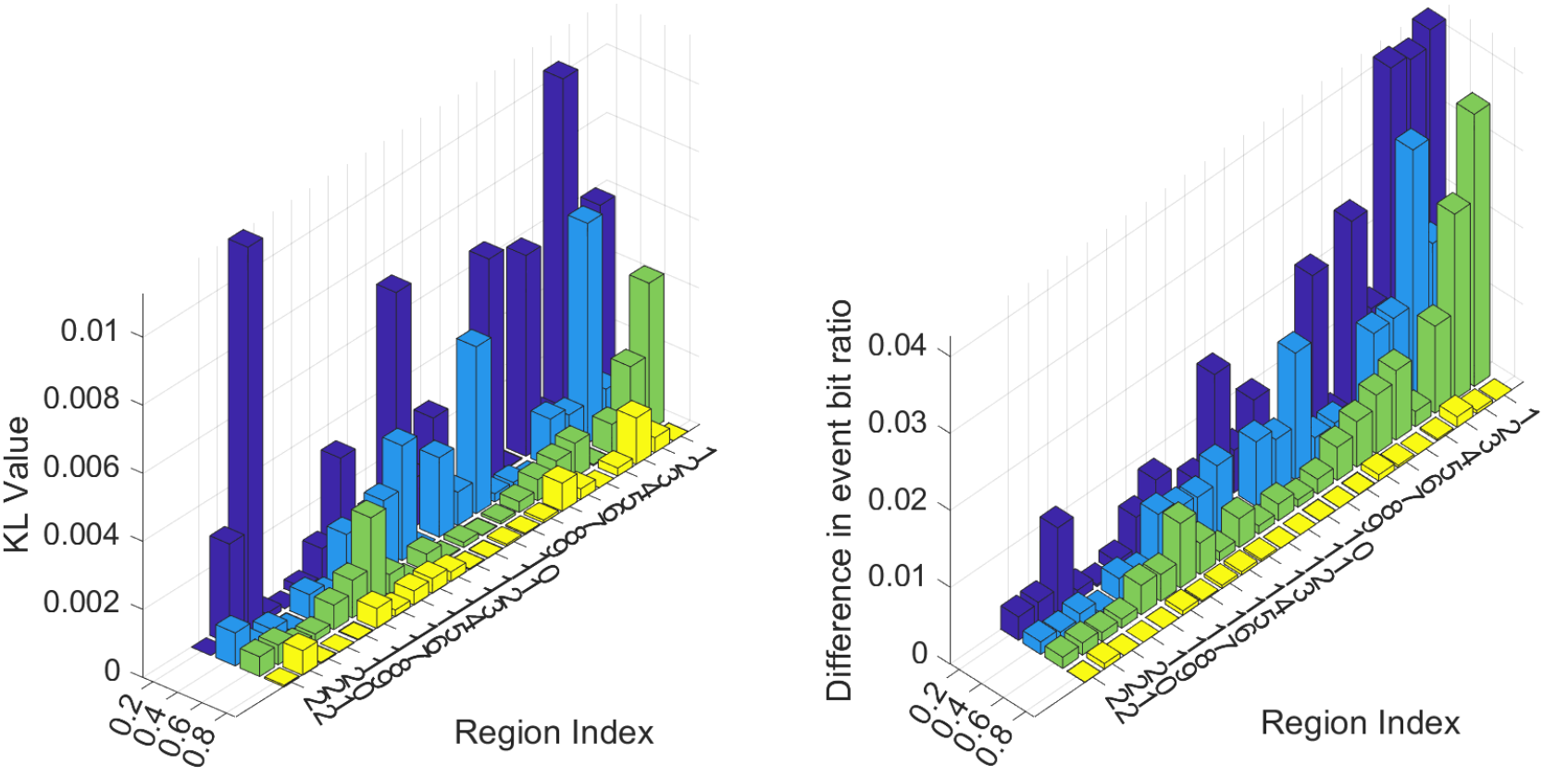}}
\vspace{-3mm}
\caption{KL divergence and event bit ratio difference between the simulator and model with fixed detection probability $P_{det}$ and varying transmission probability $P_{trans}$. Region indices correspond to Table~\ref{tab:body_parts}.}
\vspace{-4mm}
\label{fig:metrics}
\end{figure}

\begin{figure}[t]
\centerline{\includegraphics[width=0.9\columnwidth]{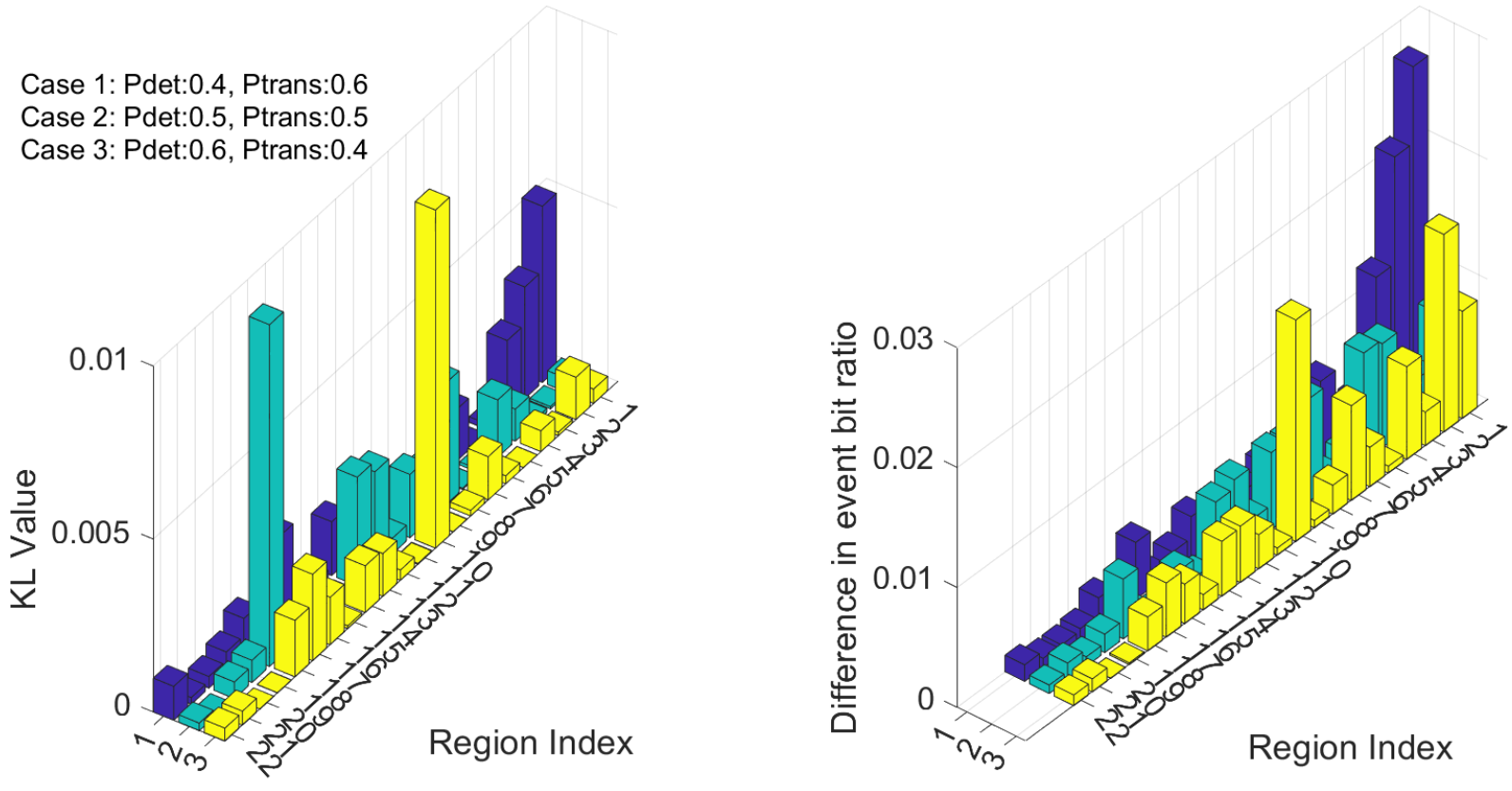}}
\vspace{-3mm}
\caption{KL divergence and event bit ratio difference between the simulator and model in mixed parameter cases. Region indices correspond to Table~\ref{tab:body_parts}.}
\label{fig:klr}
\vspace{-4mm}
\end{figure}

The \ac{KL} divergence results assessing the similarity between the event bit $b=1$ ratio in the overall datasets obtained through the simulator and the model are presented in Figure~\ref{fig:metrics}. 
As depicted, the arrangement of the regions follows a descending order of $P_{trans}$ probability. 
The results demonstrate an increase in the similarity between ratios as a function of ascending transmission probability for each region, in line with expectations. 
Conversely, a decrease in $P_{trans}$ triggers an increase in the observed divergence. 
Consequently, the \ac{KL} divergence reaches its minimum in regions with lower detection and higher transmission probability. 
Notably, the peaks in the \ac{KL} divergence emerge within less probable regions coupled with reduced transmission probabilities. 
These peaks can be attributed to the scarcity of detection of events in these regions. 
Despite the marginal discrepancies in the ratios, their diminutive nature accentuates the \ac{KL} divergence, implying a more substantial difference in such cases. 
Importantly, the observed differences do not exceed 0.04, underlining a rather high level of similarity across scenarios.

Lastly, several scenarios with mixed detection and transmission probabilities are studied to assess the model's accuracy for close-to-real-life cases.
Figure~\ref{fig:klr} depicts the outcomes for three distinct cases, while the depicted results affirm comparable performance of the model and simulator in realistic scenarios. 
However, the trends within the \ac{KL} divergence are not entirely comparable due to the presence of a few outliers in the simulator-generated raw data, indicating the presence of smaller stochastic factors beyond the model's scope.

\section{Conclusions}
\label{sec:conclusions}

In this work, we have proposed an analytical model of raw data for flow-guided in-body nanoscale localization.
Based on the nanodevices' communication and energy-related capabilities, the model outputs iteration times and event detection indicators that can be used by existing flow-guided localization approaches for localizing biological events. 
The model's output was compared with the equally-parameterized one from a simulator for flow-guided localization that features higher level of realism, indicating a significant level of similarity.

Flow-guided localization will be deployed in the cardiovascular systems of different individuals with varying biologies.
Such localization is usually based on machine learning, hence requiring significant training and corresponding data. 
This training data is hard to obtain individually, yet the tunable nature of the model could potentially be used for capturing the differences in the raw data across bloodstreams.
We consider evaluating this aspect of the model as a part of our future work, where we envision its utilization in a comparable way as for the administration of anesthesia, i.e., based on physiological indicators such as age, sex, height, and weight. 

For individual patients, temporal differences in the raw data stream are to be expected due to for example performed activities, and biological conditions (e.g., diseases) or environmental changes (e.g., temperature, humidity).
These will result in changes in the raw data stream for flow-guided localization, in turn degrading its performance. 
As a part of our future work, we envision the adaptation of the model so that it can capture these changes in individual bloodstreams, which we envisage to be used for adapting flow-guided localization based on physiological indicators such as blood pressure or heart rate.

\renewcommand{\bibfont}{\footnotesize}
\printbibliography

@article{jornet2012joint,
  title={Joint energy harvesting and communication analysis for perpetual wireless nanosensor networks in the terahertz band},
  author={Jornet, Josep Miquel and others},
  journal={IEEE Trans. on Nanotechnology},
  pages={570--580},
  year={2012},
  publisher={IEEE}
}

@inproceedings{lemic2022toward,
  title={Toward location-aware in-body terahertz nanonetworks with energy harvesting},
  author={Lemic, Filip and others},
  booktitle={ACM Nanoscale Computing and Communication},
  pages={1--6},
  year={2022}
}

@article{behboodi2016mathematical,
  title={A mathematical model for fingerprinting-based localization algorithms},
  author={Behboodi, Arash and and others},
  journal={arXiv:1610.07636},
  year={2016}
}

@article{dressler2015connecting,
  title={Connecting in-body nano communication with body area networks: Challenges and opportunities of the Internet of Nano Things},
  author={Dressler, Falko and others},
  journal={Nano Communication Networks},
  pages={29--38},
  year={2015},
  publisher={Elsevier}
}

@article{lopez2023toward,
  title={Toward Standardized Performance Evaluation of Flow-guided Nanoscale Localization},
  author={L{\'o}pez, Arnau Brosa and others},
  journal={arXiv:2303.07804},
  year={2023}
}

@inproceedings{geyer2018bloodvoyagers,
  title={BloodVoyagerS: Simulation of the work environment of medical nanobots},
  author={Geyer, Regine and Stelzner, Marc and others},
  booktitle={ACM Nanoscale Computing and Communication},
  pages={1--6},
  year={2018}
}

@article{gomez2022nanosensor,
  title={Nanosensor location estimation in the human circulatory system using machine learning},
  author={G{\'o}mez, Jorge Torres and others},
  journal={IEEE Transactions on Nanotechnology},
  volume={21},
  pages={663--673},
  year={2022},
  publisher={IEEE}
}

@article{abbasi2016nano,
  title={Nano-communication for biomedical applications: A review on the state-of-the-art from physical layers to novel networking concepts},
  author={Abbasi, Qammer H and others},
  journal={IEEE Access},
  volume={4},
  pages={3920--3935},
  year={2016},
  publisher={IEEE}
}

@article{abadal2015time,
  title={Time-domain analysis of graphene-based miniaturized antennas for ultra-short-range impulse radio communications},
  author={Abadal, Sergi and others},
  journal={IEEE Trans. on communications},
  volume={63},
  number={4},
  pages={1470--1482},
  year={2015},
  publisher={IEEE}
}

@article{lemic2021survey,
  title={Survey on terahertz nanocommunication and networking: A top-down perspective},
  author={Lemic, Filip and others},
  journal={IEEE Journal on Selected Areas in Communications},
  volume={39},
  number={6},
  pages={1506--1543},
  year={2021},
  publisher={IEEE}
}

@article{bartra2023graph,
  title={Graph Neural Network-enabled Terahertz-based Flow-guided Nanoscale Localization},
  author={Bartra, Gerard Calvo and others},
  journal={arXiv:2307.05551},
  year={2023}
}

@inproceedings{lemic2019modeling,
  title={Modeling and reducing idling energy consumption in energy harvesting terahertz nanonetworks},
  author={Lemic, Filip and others},
  booktitle={IEEE Global Communications Conference},
  pages={1--6},
  year={2019}
}

@article{lemic2023insights,
  title={Insights from the Design Space Exploration of Flow-Guided Nanoscale Localization},
  author={Lemic, Filip and others},
  journal={arXiv:2305.18493},
  year={2023}
}

@article{hossain2018terasim,
  title={TeraSim: An ns-3 extension to simulate Terahertz-band communication networks},
  author={Hossain, Zahed and Xia, Qing and Jornet, Josep Miquel},
  journal={Nano Communication Networks},
  volume={17},
  pages={36--44},
  year={2018},
  publisher={Elsevier}
}

@article{gomez2023optimizing,
  title={Optimizing Terahertz Communication Between Nanosensors in the Human Cardiovascular System and External Gateways},
  author={G{\'o}mez, Jorge Torres and others},
  journal={IEEE Communications Letters},
  year={2023},
  publisher={IEEE}
}

@article{stelzner2017function,
  title={Function centric nano-networking: Addressing nano machines in a medical application scenario},
  author={Stelzner, Marc and others},
  journal={Nano communication networks},
  volume={14},
  pages={29--39},
  year={2017},
  publisher={Elsevier}
}

\end{document}